\documentclass[runningheads, envcountsame, a4paper]{llncs}
\usepackage{graphicx}
\usepackage{amsmath,amsfonts}

\usepackage{float}
\usepackage{tikz}
\usepackage[mode=tex]{standalone}
\usetikzlibrary{shapes,arrows.meta,chains,shapes.geometric}

\usepackage{tabularx}

\usepackage{xargs}             	
\usepackage[colorinlistoftodos,prependcaption]{todonotes}
\newcommandx{\unsure}[2][1=]{\todo[linecolor=red,backgroundcolor=red!25,bordercolor=red,#1]{#2}}
\newcommandx{\change}[2][1=]{\todo[linecolor=blue,backgroundcolor=blue!25,bordercolor=blue,#1]{#2}}
\newcommandx{\info}[2][1=]{\todo[linecolor=OliveGreen,backgroundcolor=OliveGreen!25,bordercolor=OliveGreen,#1]{#2}}
\newcommandx{\improvement}[2][1=]{\todo[linecolor=Plum,backgroundcolor=Plum!25,bordercolor=Plum,#1]{#2}}
\usepackage{tikz}
\usetikzlibrary{arrows}
\makeatletter
\pgfdeclareshape{datastore}{
  \inheritsavedanchors[from=rectangle]
  \inheritanchorborder[from=rectangle]
  \inheritanchor[from=rectangle]{center}
  \inheritanchor[from=rectangle]{base}
  \inheritanchor[from=rectangle]{north}
  \inheritanchor[from=rectangle]{north east}
  \inheritanchor[from=rectangle]{east}
  \inheritanchor[from=rectangle]{south east}
  \inheritanchor[from=rectangle]{south}
  \inheritanchor[from=rectangle]{south west}
  \inheritanchor[from=rectangle]{west}
  \inheritanchor[from=rectangle]{north west}
  \backgroundpath{
    \southwest \pgf@xa=\pgf@x \pgf@ya=\pgf@y
    \northeast \pgf@xb=\pgf@x \pgf@yb=\pgf@y
    \pgfpathmoveto{\pgfpoint{\pgf@xa}{\pgf@ya}}
    \pgfpathlineto{\pgfpoint{\pgf@xb}{\pgf@ya}}
    \pgfpathmoveto{\pgfpoint{\pgf@xa}{\pgf@yb}}
    \pgfpathlineto{\pgfpoint{\pgf@xb}{\pgf@yb}}
 }
}
\makeatother

\usetikzlibrary{fit}

\tikzset{%
  >={Latex[width=2mm,length=2mm]},
    base/.style = {rectangle, rounded corners, draw=black,
                   minimum width=1cm, text centered, font=\sffamily},
    txOutBlock/.style = {base, fill=blue!30, minimum width=1cm},
    startstop/.style = {base, fill=red!30},
    repaymentBlock/.style = {diamond, rounded corners, draw=black, fill=yellow!30},
    tcMarketBlock/.style = {base, fill=green!30, minimum height=1.5cm, minimum width=4cm},
    debtCreditTxBlock/.style = {base, fill=red!30},
    process/.style = {base, minimum width=2.5cm, fill=orange!15,
                           font=\ttfamily},
    func/.style = {rounded corners, draw=black, inner sep=2pt, minimum height=2em, font=\sffamily},
}

\begin{document}

\title{Debt Representation in UTXO Blockchains\thanks{This work was supported by Mitsubishi Electric Research Laboratories.}}
\titlerunning{Debt Representation in UTXO Blockchains}
%
\author{Michael Chiu\inst{1} 
\and
Uro\v{s} Kalabi\'{c}\inst{2}
}
\authorrunning{M.~Chiu and U.~Kalabi\'{c}}
%
\institute{Department of Computer Science, University of
Toronto, ON M5S 2E4, Canada
\email{chiu@cs.toronto.edu}\\
\and
Mitsubishi Electric Research Laboratories, Cambridge, MA 02139, USA \\
\email{kalabic@merl.com}}
\maketitle              
\begin{abstract}

We provide a UTXO model of blockchain transactions that is able to represent both credit and debt on the same blockchain. Ordinarily, the UTXO model is solely used to represent credit and the representation of credit and debit together is achieved using the account model because of its support for balances. However, the UTXO model provides superior privacy, safety, and scalability when compared to the account model. In this work, we introduce a UTXO model that has the flexibility of balances with the usual benefits of the UTXO model. This model extends the conventional UTXO model, which represents credits as unmatched outputs, by representing debts as unmatched inputs. We apply our model to solving the problem of transparency in reverse mortgage markets, in which some transparency is necessary for a healthy market but complete transparency leads to adverse outcomes. Here the pseudonymous properties of the UTXO model protect the privacy of loan recipients while still allowing an aggregate view of the loan market. We present a prototype of our implementation in Tendermint and discuss the design and its benefits.

\end{abstract}

%
%

\setcounter{footnote}{0}
\section{Introduction}

There are two main blockchain transaction models: the unspent transaction output (UTXO) model and the account model. UTXOs were 
introduced in Bitcoin \cite{satoshi} 
and are predominantly used to represent credits. 
The UTXO model is stateless; UTXOs represent units of value that must be spent, so any state change within the UTXO model results in old UTXOs being succeeded by new UTXOs. The account model is state-dependent; it implements balances where transactions change the state of the system by keeping a balance \cite{antonopoulos2018eth}.
The UTXO model has various advantages over the account model including superior privacy, safety, and scalability, mainly due to the structure of transactions. However, one shortcoming 
is that it is unable to represent debt.

Debt is an important part of a well-functioning financial system \cite{holmstrom2015understanding} but its opacity within the financial system has been identified as a contributor to the 2008 financial crisis \cite{greenlaw2008leveraged}. For this reason, regulatory agencies have recommended increasing transparency in debt markets, such as mortgage markets, in order to prevent build-up of excessive leverage \cite{us_treasury_mortgage_report}. Nevertheless, 
it is understood that complete transparency also leads to adverse outcomes like increased price volatility \cite{pavlov2016transparency}. 

This work presents a possible solution to the representation of debt on UTXO blockchains. In our design, debt is represented analogously to the way in which credit is represented; where UTXOs represent credit as \emph{unspent transaction outputs}, we represent debt in something akin to \emph{unpaid transaction inputs}, i.e., a debt is a transaction that has not yet been funded. 
We implement our design in Tendermint \cite{tendermint}, a library for state machine replication (SMR), and show how it can be applied to the representation of reverse mortgage transactions on the blockchain. 
As a practical matter, we note that this work presents a protocol for managing credits and debts in an efficient manner and does not prevent debt-holders from abandoning their debt. We expect that this could be done through existing legal frameworks, or through collateralization schemes being pioneered in DeFi technology \cite{defi}. 

The literature has given some consideration to debt representation in the blockchain. One existing approach is the implementation of a debt token and logic to handle debt creation and destruction
\cite{moy2019systems,wu2020electronic}. Another approach uses the blockchain as a shared data layer to record loans \cite{dowding2018blockchain}. There have been attempts at the representation of debt in a multi-blockchain setting \cite{black2019atomic}, in which debt is represented implicitly by locking tokens on multiple blockchains and not within a single blockchain. Other work has used smart contracts to represent debt \cite{xie2020zerolender}, but it does not improve the transparency of aggregate debt within the system. Apart from work that represents debt in a blockchain, there is also at least one effort that is moving the home equity loan process onto the blockchain \cite{figure}, but it does not put mortgage transactions themselves onto the blockchain.
To our knowledge, the existing literature has not considered representing debt as unpaid transaction inputs. 

The rest of the paper is structured as follows. Section 2 discusses transactions in the UTXO model. Section 3 presents a novel way of representing debt in the UTXO model. Section 4 presents our prototype for representing reverse mortgage transactions on the blockchain. Section 5 is the conclusion.

\section{Transactions in the UTXO Model}
\label{sec:utxo}

Transactions are the fundamental data structures in blockchains that represent a state change. In the UTXO model, 
transactions are generally comprised of transaction inputs, transaction outputs, locktime and other metadata \cite{antonopoulos2014mastering}.

Transaction outputs are data structures that contain an amount, a locking script and possibly other metadata such as the size of the locking script.
Transaction amounts indicate the quantity of value to be transferred. Locking scripts within a transaction output encode the conditions that must be satisfied in order for the amount to be spent.

Transaction inputs contain a transaction hash, an output index, unlocking script, and other metadata. The transaction hash is a hash of the transaction containing the transaction output from where the value is to be drawn; transaction outputs that are not matched to a transaction input therefore are \emph{unspent}. 
The output index indicates which of the UTXOs in the referenced transaction is to be drawn from. The unlocking script contains the solution to the locking script of the referenced UTXO. 

The total amount of UTXOs represent the total amount of available credit in the system. Every node in the blockchain network keeps track of the available UTXOs in memory and this is known as a UTXO pool. When a UTXO is matched with a transaction input, it is no longer considered unspent and is removed from the UTXO pool. 

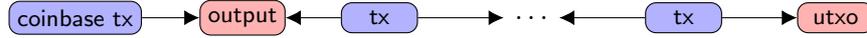
\begin{figure}[tp]
\centering
\begin{tikzpicture}
[node distance=1.5cm,
every node/.style={fill=white, font=\sffamily}, align=center]

\node (cb_tx)     at (-5,0) [txOutBlock] {coinbase tx};
\node (debtcreditTx)   at (-2.8,0) [debtCreditTxBlock] {output};
\node (tx_1)     at (-1,0) [txOutBlock] {tx};
\node (tx3)     at (1,0) [draw=none,fill=none,font=\ttfamily] {...};
\node (tx2)     at (3,0) [txOutBlock] {tx};
\node (output2)     at (5,0) [debtCreditTxBlock] {utxo};

\draw[<-] (debtcreditTx.west) -- (cb_tx.east);
\draw[<-] (debtcreditTx.east) -- (tx_1.west);
\draw[<-] (tx3.west) -- (tx_1.east);
\draw[->] (tx2.west) -- (tx3.east);
\draw[->] (tx2.east) -- (output2.west);

\end{tikzpicture}
\caption{Relationship between transactions and UTXOs in the UTXO model}
\label{fig:utxo}
\end{figure}

In permissionless UTXO blockchains, 
the coinbase transaction is used to mint new tokens in the system as rewards for miners. 
Coinbase transactions have the same fields as regular transactions and are the only transactions in a traditional UTXO model that are allowed to have unmatched transaction inputs. That is, coinbase transactions do not point to an existing UTXO. An overview of how transactions are related to each other in the UTXO model is provided in Fig.~\ref{fig:utxo}.





Although the UTXO model is well-suited to the representation of credit, it is unable to represent debt without modification. 
It is possible to modify a transaction output to hold negative values, but this would require preventing participants with permission to issue debts from sending large, negative amounts to creditors and destroying their equity. It is also possible to represent debt using smart contracts, but such an implementation  
would be opaque and require additional computation to query the amount of debt issued or owed by a debtor, and it would not allow a straightforward lookup on the blockchain by parsing transactions.
Another possible change that could be made to the UTXO model is to simply double the data fields of the transaction so that the duplicate set of fields represent debt tokens. However, not only does this double the size of transactions, it also allows users to send their debts away.
We introduce what we believe to be a more elegant solution, detailed in the following.

%
%

\section{Debt-Enabling UTXO Blockchain}
\label{sec:utxi}

We present a design that enables the representation of debt in a UTXO-based permissioned blockchain. 
In the design, we represent debts as transactions with unmatched transaction inputs, conforming to the way in which 
credits are represented using unmatched outputs. 
To enable this representation, we introduce two new types of transactions: debt transactions and outstanding debt transactions.



%
%
\subsection{Debt Transactions}
The first type of transaction is a \emph{debt transaction}. These transactions enable debt to be issued from creditor to debtor. They are similar to coinbase transactions in that they are transactions with unmatched transaction inputs and are only constructed by a subset of network participants. These participants must be given permission to issue debt,
and the permissioning model itself 
can vary. For example, it can be 
that every trusted participant is able to issue debt, such as in a permissioned network of banks, or that only a subset of trusted participants is able to do so, 
such as in the case of a 
single central bank. 
To this end, special addresses known to the permissioned network protocol, and their corresponding public and private keys, are used to confer the ability to issue debt. 


Although similar to coinbase transactions, debt transactions are markedly different in that the transaction input has an actual function. 
In a debt transaction input, the erstwhile   
transaction hash field is re-purposed to act as a public key field that records a public key belonging to the creditor; 
this enables parties involved in the debt issuance to be recorded on the blockchain. The output index is also repurposed and set to, for example, $-2$ in order to provide a simple flag to check if the transaction is a debt transaction. 

The rest is similar to a coinbase transaction in that  
the debt transaction output is a normal transaction output. 
Since the transaction output of the debt transaction is a standard UTXO, it is included into the UTXO pool like conventional UTXOs after the debt transaction has been accepted by the network.


\begin{figure}[tp]
\centering
\begin{tikzpicture}
[node distance=1.5cm,
every node/.style={fill=white, font=\sffamily}, align=center]
\node (cb_tx)     at (0,0) [txOutBlock] {debt tx};
\node (tx_1)     at (3,0) [debtCreditTxBlock] {utxo};

\draw[<-] (tx_1.west) -- (cb_tx.east);

\node (outstand1) at (0,-0.8) [txOutBlock] {outstanding debt tx};
\node (output3)   at (-3,-0.8) [debtCreditTxBlock] {output};

\draw[dashed] (outstand1.north) -- (cb_tx.south);   
\draw[->] (output3.east) -- (outstand1.west);

\end{tikzpicture}
\caption{Debt creation on the blockchain}
\label{fig:debt_tx}
\end{figure}
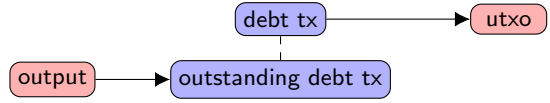


The UTXO nature of debt transactions protect the privacy of the debtor since the debt issuance can be split across multiple address, i.e., transaction outputs. The aggregate amount of issued debt remains public and is recorded directly on the blockchain, increasing transparency into the health of the overall system while protecting individual privacy.

%
%
\subsection{Outstanding Debt Transactions and Debt Pools}

Debt transactions are broadcast to the network for inclusion into the blockchain and create UTXOs assigned to the debtor. Debt transactions act as mechanisms to issue debt and are recorded on the blockchain. However, since one needs a mechanism to keep track of outstanding debts and their repayment, we introduce \emph{outstanding debt transactions}, which are created simultaneously with debt transactions, as shown in Fig.~\ref{fig:debt_tx}. Outstanding debt transactions are transactions with unmatched inputs and a transaction output 
matched with the corresponding debt transaction and the creditor's public key.

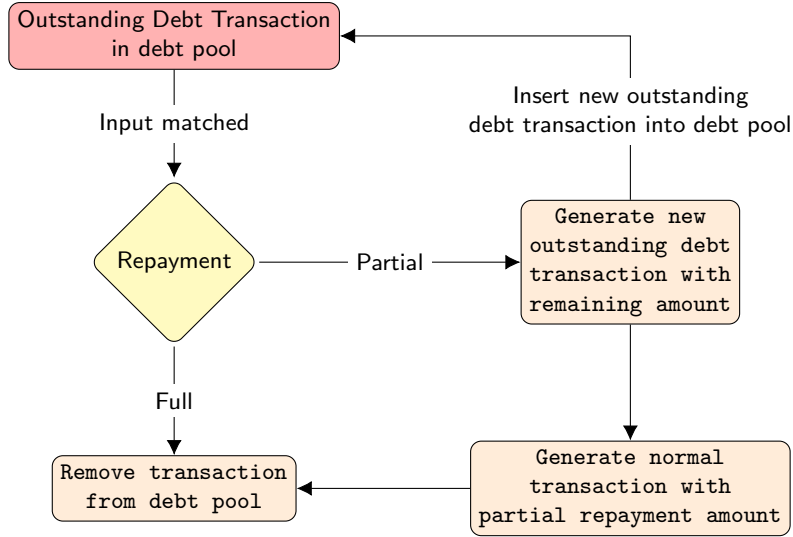
\begin{figure}[t]
\centering
\begin{tikzpicture}[node distance=1.5cm,
every node/.style={fill=white, font=\sffamily}, align=center]

\node (fullBlock)       at (0,0) [process] {Remove transaction \\ from debt pool};   
\node (partialBlock)    at (6,3) [process] {Generate new \\ outstanding debt \\ transaction with \\ remaining amount};   
\node (removeBlock)    at (6,0) [process] {Generate normal \\ transaction with \\ partial repayment amount}; 
\node (debtBlock)       at (0,6) [startstop] {Outstanding Debt Transaction \\ in debt pool};
\node (repayBlock)      at (0,3) [repaymentBlock] {Repayment};

\draw[->]       (debtBlock) --  node{Input matched} 
                                (repayBlock);
                                
\draw[->]       (repayBlock) -- node{Full}
                                (fullBlock);
                                
\draw[->]       (repayBlock) -- node{Partial}
                                (partialBlock);
                                
\draw[->]       (partialBlock) |- node[yshift=-1cm]
                                  {Insert new outstanding \\ debt transaction into debt pool}
                                  (debtBlock);
                                 
\draw[->]       (partialBlock) -- (removeBlock);
\draw[->]       (removeBlock) -- (fullBlock);

                                        



\end{tikzpicture}
\caption{Lifecycle of an outstanding debt transaction in the debt pool}
\label{fig:outstanding_debt_tx_repayment}
\end{figure}

After an outstanding debt transaction is created and broadcast to all other nodes, it is inserted into a \emph{debt pool}, which is similar to a UTXO pool, but which holds outstanding debt transactions instead of UTXOs. The debt pool is used to handle debt repayments. An outstanding debt transaction is removed from the debt pool when a debt owner repays the remainder of a debt, i.e., when the outstanding debt transaction's input is matched with a debtor's UTXO containing the funds. Once an outstanding debt transaction has a matched input, it is no different from a normal transaction 
and is removed from the debt pool and inserted into the transaction pool to be eventually accepted by the network.
In the case of partial repayments, 
we create two new transactions from the original outstanding debt transaction: the first is a normal transaction that records the transfer of value of the repayment amount from the debtor to the creditor; 
the second 
is a new outstanding debt transaction with the remaining debt amount, which is similar to the way change transactions are handled in UTXO blockchains. The issuance of two new transactions in the case of repayments ensures that funds allocated to repayment and outstanding debt balances are finalized by the network and not held in debt pools, akin to splitting a UTXO when a transfer of credit is made. Fig.~\ref{fig:outstanding_debt_tx_repayment} illustrates the lifecycle of an outstanding debt transaction.


%
%



\section{Prototype}
\label{sec:prototype}
We implement a prototype of a UTXO-based blockchain capable of debt representation and apply it to reverse mortgages.\footnote{For code listing, see \url{https://github.com/chiumichael/debtchain}} Reverse mortgages, also known as home equity loans, allow home owners to 
use their primary residence as collateral for a loan. 
The reverse mortgage market is an important source of wealth for many households and borrowing against home equity is a significant percentage of US household leverage \cite{mian2011house}. As a significant contributor to the over-leveraging of many households, reverse mortgages are potentially a large source of systemic risk in the financial system and 
one of the contributing factors to the financial crisis of 2008 \cite{greenlaw2008leveraged}. For this reason, there have been many recommendations by regulatory bodies since 2008 that recommend 
increased transparency in mortgage markets \cite{us_treasury_mortgage_report}. 
However, transparency in the current system is difficult because these transactions are private and the relevant data is siloed 
\cite{alter2020non}.

A permissioned UTXO blockchain for mortgage transactions offers a solution amenable to all participants since it enables transparency at a system level while preserving individual privacy. Note that an account-based blockchain is unsuitable for this because, unlike UTXO-based transactions, balance-based transactions are unable to provide transaction-level privacy protections for loan recipients. 
To protect privacy, a loan issuance implemented using a debt transaction can have multiple UTXOs, obfuscating both the possible number of recipients and the total amount issued per recipient. This is in contrast to the account model, where balanced-based transactions reinforce the reuse of balances because they are state-dependent.

\subsection{System Architecture}
\begin{figure}[tp]
\centering
\begin{tikzpicture}[
  font=\sffamily,
  every matrix/.style={ampersand replacement=\&,column sep=2cm,row sep=2cm},
  source/.style={draw,thick,rounded corners,fill=green!20,inner sep=.3cm},
  process/.style={draw,thick,circle,fill=blue!20},
  client/.style={draw,thick,circle,fill=yellow!20},
  sink/.style={source,fill=green!20},
  datastore/.style={draw,very thick,shape=datastore,inner sep=.3cm},
  dots/.style={gray,scale=2},
  to/.style={->,>=stealth',shorten >=1pt,semithick,font=\sffamily\footnotesize},
  every node/.style={align=center}]

\node[source] (hisparcbox) at (3,-1) {UTXO pool};
\node[datastore] (storage) at (-3,0) {Tendermint \\ Engine};
\node[process] (monitor) at (0,0) {node};
\node[sink] (datastore) at (3,1) {Debt Pool};

\node[label={[name=l]}, client] at (0,-2.4) (client) {client};


\draw[<->] (hisparcbox) -- (monitor);

\draw[to] (monitor) to[bend right=50] node[midway,above] {block hashes}
  node[midway,below] {\texttt{abci}} (storage);
\draw[to] (storage) to[bend right=50] node[midway,above] {\texttt{abci}}
  node[midway,below,label={[name=j, yshift=-0.2cm]}] {block hashes} (monitor);
  
\draw[<->] (client) -- (monitor);
\draw[<->] (monitor) -- (datastore);

\tikzset{blue dotted/.style={draw=blue!50!white, line width=2pt,
                           dash pattern=on 1pt off 4pt on 6pt off 4pt,
                            inner sep=4mm, rectangle, rounded corners}};

\tikzset{red dotted/.style={draw=red!50!white, line width=1pt,
                       dash pattern=on 1pt off 4pt on 6pt off 4pt,
                        inner sep=4mm, rectangle, rounded corners}};


\node (first dotted box) [blue dotted,
                        fit = (storage) (hisparcbox) (monitor) (datastore) (j)] {};

\node (second dotted box) [red dotted,
                        fit = (hisparcbox) (monitor) (datastore)] {};
                            
\end{tikzpicture}
\caption{System level diagram for the prototype, where the blue box contains the back-end components and the red box contains the in-node components} 
\label{fig:system_diargram_home_equity}
\end{figure}
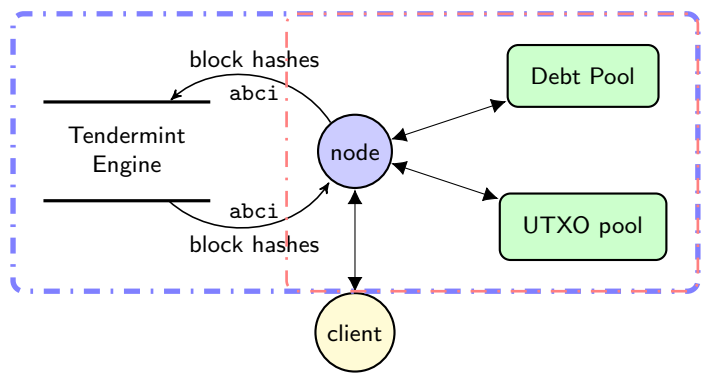

Our blockchain prototype is built on top of Tendermint Core\footnote{\url{https://github.com/tendermint/tendermint}} \cite{tendermint}, an open-source Byzantine fault tolerant (BFT) middleware library for SMR. 
Tendermint Core 
provides a consensus mechanism 
for  both permission\textbf{}ed and permissionless BFT blockchains 
and includes 
the rotation of the leader after every round, using gossip protocols to communicate with other nodes. 
Tendermint is comprised of two main components: the Tendermint engine and the ABCI, which is the Tendermint API. 
The engine handles the consensus and the dissemination of information throughout the network and 
the ABCI provides an interface for an application to interact with the consensus mechanism. The main benefit of separation of consensus from the application logic is that it allows applications to only consider the local state, not having to explicitly manage synchronization.
See Fig.~\ref{fig:system_diargram_home_equity} for a schematic.

The first component of our prototype is \texttt{pkg/utxi}, a modular library that implements UTXO transactions, along with the debt and outstanding-debt transactions introduced in this work. 

The second component implements the blockchain functionality: UTXO and debt pools, and block construction for network acceptance. The memory pools are implemented as in-memory key-value stores using BadgerDB\footnote{\url{https://github.com/dgraph-io/badger}} at the application level; they are implemented on top of the ABCI since they rely on the ABCI to receive instructions. Blocks are constructed implicitly; the Merkle root of the hash of the transactions in a block is broadcast to the network for acceptance. 
Transactions, within blocks, are stored internally in BadgerDB 
on top of the ABCI. Tendermint handles the propagation of state changes made by transactions.

The final component is the application node and the client. The application node contains the functionality necessary for a reverse mortgage blockchain such as issuing debt, checking balances, checking the amount of existing debt, among other things, and is implemented on top of the ABCI at the same level of blockchain functionality. The client is a front-end for users to interact with the network and contains user-related functionality such as private key management. There can be many clients connected to a single node, 
which communicate with the node through HTTP requests.

\subsection{Implementation}

Home equity loans are implemented as debt transactions. The input of the debt transaction's public key field is interpreted as the issuer 
and its data field indicates the type of loan.  

The backend (Fig.~\ref{fig:system_diargram_home_equity}; blue box) receives commands from users, through the client (Fig.~\ref{fig:system_diargram_home_equity}; yellow circle) and the ABCI interface. 
The two main ABCI interface functions that must be implemented are \texttt{CheckTx} and \texttt{DeliverTx}, which are entry points into the Tendermint engine. Logic for checking the validity of requests is implemented in \texttt{CheckTx}. Logic that changes the state of the application is implemented within or called only from \texttt{DeliverTx}. Since Tendermint handles consensus and the replication of state among nodes, it needs to be able to invoke functions changing state.

When a request to issue debt is sent from a client, a debt transaction is constructed and is broadcast to the network through the \texttt{DeliverTx} function. Simultaneously, the local state of the node is updated and the UTXO pool is updated with the UTXOs belonging to the debtor, while the debt pool is updated with an outstanding debt transaction belonging to the creditor.

%
%

%
%
\section{Conclusion}

In this short paper, we introduced a permissioned, UTXO-based blockchain design that is able to represent credit and debt on the same blockchain. The main idea behind the design is that debt can be represented as unmatched transaction inputs in the same way that credits are represented as unmatched transaction outputs. To handle this new construction, we introduced a debt pool, similar to a UTXO pool, that keeps track of debts in the system.
A benefit this provides is the ability to keep track of aggregate debt while protecting 
individual privacy, because one loan can be represented with multiple, pseudonymous UTXOs. We presented a prototype of our design applied to the representation of reverse mortgages. 






\bibliographystyle{splncs04unsrt} 
\bibliography{main}

\end{document}